\documentclass[preprint,aps]{revtex4}

\usepackage{epsfig}

\begin{document}

\title{Influence of progressive central hypovolemia \\on  multifractal dimension of cardiac interbeat intervals.}
\author{Bruce J. West$^{1,2,3}$, Nicola Scafetta$^{1,2}$, William H. Cooke$^4$ and Rita
Balocchi$^5$}
\address{$^{1}$ Physics Department, Duke University, Durham, NC 27708}
\address{$^{2}$Pratt School EE Dept., Duke University,  P.O. Box 90291, Durham, NC 27708 }  
\address{$^{3}$ Mathematics Division, Army Research Office, Research Triangle Park, NC 27709. } 
\address{$^4$ Departments of Biomedical Engineering and Biological Sciences, Michigan Technological University, Houghton MI 49931.}
\address{$^5$ Istituto Fisiologia Clinica, CNR-Area Ricerca San Cataldo, Via Moruzzi 1, 56124-Pisa}

\date{\today}

\begin{abstract}
We analyzed the heartbeat time series of 12 human subjects exposed to progressive central hypovolemia with lower body negative pressure. Two data processing techniques based on wavelet transforms were used to determine the change in the non-stationary nature of the time series with changing negative pressure. Our results suggest that autonomic neural mechanisms driving cardiac interbeat intervals during central hypovolemia go through various levels of  multifractility.
\end{abstract}

\maketitle


\section{Introduction}

In the past two decades we have witnessed an explosion in the biophysics and
physiological literature with regard to the identification of phenomena having long-term memory and probability densities that extend far beyond the typical tail region of Gaussian distributions. One way these processes have
been classified is as 1/f-phenomena, since their time series have spectra
that are inverse power law in frequency or their probabilities have an inverse power-law distribution. In either case the underlying structure is fractal,
either in space, time or both \cite{bass94}. Physiological signals, such as
the heartbeat interval time series studied herein, are typically generated by complex self-regulatory systems that process inputs with a broad range of characteristics. Such physiological time series fluctuate in an irregular
and complex manner such as shown in Fig. 1 and the statistics of these
fluctuations often exhibit self-affine or fractal properties \cite{bass94}.
Such processes do not possess dominant time scales, but they can frequently
be characterized by  fractal dimensions \cite{feders}. Ivanov \textit{et al} 
\cite{ivanov99} established that healthy human heartbeat intervals, rather
than being fractal, exhibit multifractal properties and they uncovered the loss
of multifractality for a life-threatening condition known as congestive
heart failure. Latka \textit{et al.} \cite{latka02} similarly determined
that cerebral blood flow in healthy humans is also multifractal.

Recently, Scafetta \textit{et al.} \cite{scafetta1} studied how the fractal
and multifractal nature of the human stride interval is modified by changing
the pace velocity in both free and metronomically constrained walking
conditions and introduced a neuronal super central pattern generator (SCPG) to
model the human stride interval time series \cite{scafetta2}. In particular,
it was determined that an increase of neural stress, induced by a faster or slower pace than normal, causes a slight increase of both fractal and multifractal properties of the free pace stride interval time series \cite{scafetta1,scafetta2}. In the present analysis we used lower body negative pressure (LBNP) to induce progressive central hypovolemia and determined how such stimulation of autonomic neural control mechanisms modified the statistical nature of heart-beat time series. A LBNP is applied by sealing supine subjects at the level of the iliac crest in an air-tight chamber connected to a vacuum pressure control system. The magnitude of LBNP is carefully regulated to induce progressive central hypovolemia and autonomic cardiovascular compensatory mechanisms are evaluated. LBNP decreases central venous pressure, stroke volume and cardiac output by inducing a footward fluid shift, and increases heart rate and total peripheral resistance through parasympathetic neural withdrawal and sympathetic neural activation \cite{Covertino}. The magnitude of changes in autonomic neural activity can be estimated non-invasively in the frequency domain, but common analyses incorporating either fast-Fourier based power spectra \cite{Akselrod} or autoregressive modeling \cite{Malliani}  do not account for inherent multifractality or non-stationarity underlying complex cardiovascular regulatory mechanisms. For this reason, in the present study we applied two wavelet-based techniques to study the fractal and multifractal dependence of the cardiovascular responses to progressive central hypovolemia to determine how cardiovascular stress affects fluctuations of the data on a hierarchy of time scales.  For example, the
analysis allows us to isolate the influence of the breathing cycle component
on heart rate. We determine that in this situation, similar to the situation
in gait dynamics \cite{scafetta1,scafetta2}, an induced stress, in this case through central hypovolemia, causes a slight increase in both the
fractal and multifractal properties of the heartbeat interval time series.
We suggest that such changes in fractal and multifractal properties characterize neural network responses of biological systems to physiological stress.

In Sec. II we present the heart interbeat interval data analyzed herein; Sec. III is devoted to a review of the techniques of analysis; Sec. IV is devoted to a detailed analysis of the heartbeat interval time series of 12 normal subjects and in Sec. V we draw some conclusions.

\section{Heart interbeat data }
In the main experiment we studied twelve healthy men between the ages of 26 and 40 after they had been informed of the nature of the experiment and had signed a consent form, approved by the institutional human use committee of the U.S. Army Institute of Surgical Research, Ft Sam Houston, TX. Subjects were positioned supine in an LBNP chamber, sealed at the waist to the level of the iliac crest, and instrumented with a standard 4-lead electrocardiogram. Because respiratory frequency and depth may seriously confound interpretation of autonomic rhythms \cite{Brown}, subjects breathed in synchrony with a metronome set at a pace of 15 breaths per minute (0.25 Hz). Electrocardiogram data were sampled at 500 Hz during 3-minute periods of controlled frequency breathing at negative pressure of 0, -15, -30, -45, -60, and again at 0 mmHg. Data were recorded directly to computer (Windaq, Dataq Instruments, Akron, OH) and then imported into a customized software program for analysis [Cardio-Pulmonary Research Software for Windows, (WinCPRS), Absolute Aliens Ay, Turku, Finland]. R-waves were automatically detected from continuous waveforms and marked at their occurrences in time to obtain heartbeat periods (tachogram) as shown in Fig. 1.     

In a second experiment, conducted at the Institute of Clinical Physiology, CNR-Pisa, we compared the fluctuations of a tachogram with the contemporary respiratory signal regarding one person. This experiment was originally done with thirteen healthy volunteers. These volunteers underwent an experimental session in which an electrocardiographic (ECG) derivation and the respiratory
signal were simultaneously and continuously recorded at 1000 Hz
sampling frequency. The ECG was recorded using standard
electrodes, while the respiratory signal was detected through a
polymeric piezoelectric dc-coupled transducer inserted into a belt
wrapped around the chest. During the experimental session the
subjects were comfortably sitting and breathing freely.

\section{Fractal and multiresolution analysis based on wavelets}

The analysis we apply to heartbeat time series emphasizes the scaling nature
of the phenomenon. In the next two subsections we briefly review the two
techniques we use in our analysis. The first method determines the
distribution of H\"{o}lder exponents of a time series, by means of the
continuous wavelet transform (CWT) \cite{struzik}. This technique is used to study the
fractal and multifractal nature of a time series and has already been used  to analyze the human stride interval time series \cite{scafetta1}.  The second method uses the
wavelet multiresolution analysis (WMA) of a time series, by implementing the
maximum overlap discrete wavelet transform (MODWT) \cite{percival}. The latter technique
provides a detailed study of the fluctuations of a time series on a \textit{%
scale-by-scale} basis.

\subsection{Approximate estimation of local H\"older exponents and their
probability distribution.}

Mandelbrot \cite{2Mandelbrot} showed that many natural phenomena are described by self-affine,
correlated, scaling time series and as we mentioned above, heartbeat data
fall into this category. The scaling properties of the fractal noise studied
by Mandelbort, fractional Gaussian noises (fGn) \cite{2Mandelbrot}, are characterized by an
exponent that he called $H$ in honor of Hurst. Consequently, if $X(t)$ is a
fractal process with Hurst exponent $H$ and $c$ is a constant, then $X\left(
t\right) =X(ct)/c^{H}$ is another fractal process with the same statistics.
The phenomenon of $fGn$ has a spectrum with the inverse power-law form 
\begin{equation}
P(f)\propto f^{-\beta }=f^{1-2H}\approx f^{-1-2h_{0}}~,  \label{spect}
\end{equation}
where $f$ is the frequency, $H$ is the Hurst exponent and $h_{0}$ is the
average of the H\"{o}lder exponent distribution among the singularities of
the time series. We can also express a fractal noise $\{\xi_i\}$ through the equivalent representation of the autocorrelation function \cite{politi} 
\begin{equation}
C(r)=\frac{\langle \xi_{i}~\xi_{i+r}\rangle}{\langle \xi_{i}^{2} \rangle}\propto r^{2H-2} \approx r^{2h_{0}}~.  \label{cor111}
\end{equation}

The self-affine property expressed by (\ref{spect}) and the
relation between $\beta $, $H$ and $h_{0}$ are theoretically valid only for
a infinitely long monofractal time series. Figure 2 shows a computer
generated realization of $fGn$ with Hurst exponent $H=1$ or H\"{o}lder
exponent $h_{0}\approx 0$, also known as $1/f$ noise or \textit{pink} noise.
This type of noise is important because it represents a threshold between
the persistent-stationary noise ($-0.5<h_{0}<0$) and the non-stationary
noise ($h_{0}>0$). Moreover, the $1/f$ noise is characterized by an autocorrelation function $C(r)$ that does not depend on the distance $r$, Eq. (\ref{cor111}). 
The \textit{random noise} or 
\textit{white noise} is characterized by $h_{0}\approx -0.5$ whereas the 
\textit{random walk} or \textit{Brownian motion} is characterized by $%
h_{0}\approx 0.5$. In general, the higher the H\"{o}lder exponent, the
smoother the time series.

As seen in Figure 2, a fractal noise is characterized by trends and
discontinuities that give a particular geometric shape to the signal. The
rapid changes in the time series are called \textit{singularities} of the
signal and their strength is measured by a H\"{o}lder exponent \cite{struzik}%
. Given a function $f(x)$ with a singularity at $x_{0}$, the H\"{o}lder
exponent $h(x_{0})$ at such a point is defined as the supremum of all
exponents $h$ that fulfills the condition: 
\begin{equation}
|f(x)-P_{n}(x-x_{0})|\leq C|x-x_{0}|^{h}~,  \label{holdedef}
\end{equation}
where $P_{n}(x-x_{0})$ is a polynomial of degree $n<h$.

The H\"{o}lder exponent of a singularity can be evaluated by using the
wavelet transform \cite{daubechies,Mallat,percival}. Wavelet transforms make
use of scaling functions that have the property of being localized in both
time and frequency. A scaling coefficient $s$ characterizes and measures the
width of a wavelet. Given a signal $f(x)$, the continuous wavelet transform
(CWT) of $f(x)$ is defined by 
\begin{equation}
W_{s,x_{0}}(f)=\int\limits_{-\infty }^{\infty }\frac{1}{s}~\psi \left( \frac{%
x-x_{0}}{s}\right) ~f(x)~dx~,  \label{cwtdhjj3}
\end{equation}
where the kernel $\psi (u)$ is the wavelet filter centered at the origin, $%
u=0$, with unit width, $s=1$. A widely adopted choice of the kernel 
$\psi (u)$ is the so-called Mexican Hat, that is, the second derivative of a
Gaussian.

The wavelet transform can be used to determine the H\"older exponent of a
singularity because the wavelet kernel $\psi(u)$ can be chosen in such a way
as to be orthogonal to polynomials up to degree $n$, that is, such that the
following properties are fulfilled 
\begin{equation}  \label{wavort}
\int\limits_{-\infty }^{+\infty } \frac{1}{s}~ \psi\left(\frac{x-x_0}{s}
\right)~x^m~dx=0~~~\forall m, ~0\leq m \leq n~.
\end{equation}
In fact, if (\ref{wavort}) holds true, it is easy to prove that if the
function $f(x)$ fulfils condition (\ref{holdedef}), its wavelet transform at $%
x=x_0$ is given by 
\begin{equation}  \label{wavcoeff}
W_{s, x_0}(f)=C |s|^{h(x_0)} \int\limits_{-\infty }^{+\infty
}|u|^{h(x_0)}~\psi(u)~du \propto |s|^{h(x_0)}~,
\end{equation}
where $u=(x-x_0)/s$. Therefore, at least theoretically, the H\"older
exponent of a singularity that is localized in $x_0$, can be evaluated as
the scaling exponent of the wavelet transform coefficient, $W_{s, x_0}(f)$,
for $s \to 0$.

Even if Eq. (\ref{wavcoeff}) can be evaluated for any position $x_0$, we are
interested only in the cusp singularities of the time series. Mallat et al. 
\cite{Mallat,mallat2,mallat3} show that the H\"older exponent of these
singularities can be evaluated by studying the scaling exponent $h(x_0)$
along the so-called \textit{maxima line} that converges towards the
singularity. The maxima lines are defined by the extremes of the wavelet
transform coefficients (\ref{cwtdhjj3}) at each wavelet scale $s$. Arneodo
et. al. \cite{arneodo1,arneodo2} proved that WTMM can be used to define a
multifractal-like formalism that gives the stochastic properties of the
singularities of a fractal or multifractal noise. This methodology has been
recently used to determine the multifractal nature of many signals, for
example, that for human heartbeats \cite{ivanov99}.

However, the above method presents some problems of stability when applied
to observational data. In fact, the spectrum can be corrupted by the
divergences of negative moments \cite{Mallat,arneodo2,struzik2} or by the 
\textit{outliers}, that is, the end points of the sample singularities \cite
{struzik}. Different methods have been suggested to remove the divergences
due to the negative moments of the multifractal partition function, for
example, by chaining the wavelet maxima across scales \cite{Mallat} or, more
efficiently, by bounding the H\"older exponent of the maxima line by using
the \textit{slope} wavelet \cite{struzik2}. The corruption of the
singularity spectrum due to the outliers is more difficult to deal with.

Struzik \cite{struzik} suggested an alternative method that has the ability
to determine an approximate value of local singularity strength. The
spectrum may then be evaluated from these approximate values. The idea is to
estimate the mean H\"{o}lder exponent $\overline{h}$ as a linear fit of the
following equation 
\begin{equation}
\log [M(s)]=\overline{h}~\log (s)+C~,  \label{linfitesa}
\end{equation}
where the function $M(s)$ is obtained via the partition function, and the
mean H\"{o}lder exponent $\overline{h}$ is the local version of the Hurst
exponent $H$ \cite{2Mandelbrot}. More precisely, for a monofractal noise
with Hurst exponent $H$, we have $\overline{h}=H-1$ because the Hurst
exponent is evaluated by integrating the noise \cite{2Mandelbrot,feders}.
Here again the equality is only rigorously true for an infinitely long
monofractal noise data set. The approximate local H\"{o}lder exponent $\hat{h%
}(x_{0},s)$ at the singularity $x_{0}$ can now be evaluated as the slope 
\begin{equation}
\hat{h}(x_{0},s)=\frac{\log (|W_{s,x_{0}}(f)|)-(\overline{h}~\log (s)+C)}{%
\log (s)-\log (s_{N})}~,  \label{locholexp}
\end{equation}
where $s_{N}$ is the length of the entire wavelet maxima line tree, that is,
the maximum available scale that coincides with the sample length $s_{N}=N$,
and $x_{0}$ belongs to the set $\Omega (s)$ of all wavelet maxima at the
scale $s$ that assume the value $W_{s,x_{0}}(f)$.

While further details may be found in the papers by Struzik \cite{struzik}
and in Ref. \cite{scafetta1}, here we focus on the
interpretation of the final output, that is, the distribution of estimated
local H\"{o}lder exponents of a time series as  given in \cite
{scafetta1}. Fig. 3 shows the histogram of H\"{o}lder exponents obtained
using a computer-generated data set with a Hurst exponent $H=1$ shown in
Fig. 2. The histogram is fitted with a normalized Gaussian function of the
type 
\begin{equation}
g(h)=\frac{1}{\sqrt{2\pi }~\sigma }~\exp {\left[ -\frac{(h-h_{0})^{2}}{%
2~\sigma ^{2}}\right] }  \label{gaussf}
\end{equation}
that captures the two main characteristics of the time series: (a) the
average of H\"{o}lder exponent $h_{0}$ that gives an estimate of the
fractal nature of a time series; (b) the width of the distribution given by
the standard deviation $\sigma $ that measures the variability of the local
H\"{o}lder exponents, that is, the multifractality of a time series.
Finally, as Scafetta \textit{et al.} \cite{scafetta1} explained, a
monofractal time series presents a particular width of the H\"{o}lder
exponent distribution that depends on the length of the time series, as Eq. (%
\ref{locholexp}) suggests. Therefore, a time series will present a
multifractal nature only if the width of its H\"{o}lder exponent
distribution is larger than that of a monofractal time series of the same
length. In general, given two time series of the same length, if $\sigma
_{1}>\sigma _{2}$, we may say that the time series `$1$' has a stronger
multifractal character than time series `$2$'.

\subsection{Wavelet multiresolution analysis.}

The second wavelet-based technique that we implement allows a detailed study of
the fluctuations of a time series on a \textit{scale-by-scale} temporal
basis. The MODWT \cite{percival} is almost independent of the particular
family of wavelets used in the analysis and is the basic tool needed for
studying the multiresolution analysis of time series of $N$ data points via
wavelets. In the book of Percival and Walden, \cite{percival}, the reader can
find all the mathematical details.

In this paper we make use of the Daubechies \textit{least asymmetric}
scaling wavelet filter (LA8) that looks like the Mexican Hat, but  is also
weakly asymmetric; a fact that makes LA8 filter more malleable than the
Mexican Hat. The WMA via MODWT establishes that given an integer $J$ such
that $2^{J}<N$, where $N$ is the number of data points, the original time
series represented by the vector $\mathbf{X}$ can be decomposed on a
hierarchy of time scales represented by a smooth part plus details as
follows: 
\begin{equation}  \label{decomrel}
\mathbf{X}=S_{J} + \sum _{j=1}^{J} D_j~,
\end{equation}
with the quantity $S_{j}$ generated by the recursion relation 
\begin{equation}  \label{decomrelrr}
S_{j-1}= S_{j} + D_{j}~.
\end{equation}
The detail $D_j$ of Eq. (\ref{decomrel}) represents changes on a scale of $%
\tau=2^{j}$, while the smooth $S_{J}$ represents the smooth wavelet averages
on a scale of $\tau_{J}= 2^{J}$. In the same way we refer to as {\it residuals} the curves
\begin{equation}  \label{decomw}
R_{J} = \sum _{j=1}^{J} D_j~,
\end{equation}
such that $\mathbf{X}=S_{J}+R_{J} $.

Fig. 4 shows that WMA of the heartbeat data shown in Fig. 1 for J=4. The
detail curves show the main characteristic fluctuation that characterize
each time scale that may be expressed in the number of beats, as done in Fig. 4,
or in physical time units. We shall use these detail curves to establish the
dependence on the pressure of the period measured in both number of beats
and in physical time units of the  fluctuations of the heartbeat at
different wavelet scale.

\section{Analysis of the heartbeat time series}

Fig. 1 shows the heartbeat time series at the four  conditions of LBNP for a
typical individual. In this work, we analyze the heartbeat interval time
series for 12 individuals. In the time series data shown in the figure the increase in the average
heart rate with increasing negative pressure is clear,
as is the return to normal with the return of the negative pressure to zero.
To emphasize this effect, in Fig. 5 we plot  the average heartbeat
interval  against the pressure for 12 individuals. The figure shows that the
average heartbeat interval manifests a steady decrease from $\overline{\Delta t}%
\approx 980\pm 50$ msec at $P=0$ mmHg to $\overline{\Delta t}\approx 710\pm
50$ msec at $P=-60$ mmHg. At the recovery, the average heartbeat interval is
slightly larger than the initial one; we measure $\overline{\Delta t}\approx
1050\pm 50$ msec. The error bars measure the average standard deviation of the
variability of the heartbeat interval for each individual and do not seem to
significantly change with the pressure.

A careful visual analysis of Fig. 1 suggests that by increasing the negative
pressure, the heartbeat interval time series  becomes smoother. As explained in the previous section, a decrease of
randomness or an increase of the smoothness of a time series may be easily
detected by the H\"{o}lder exponents of the time series that measure the
strength of the singularities. The H\"{o}lder exponents increase with the
smoothness of a time series. Fig. 6 shows an estimation of the H\"{o}lder
exponents of the time series of the heartbeat interval time series shown in
Fig. 1. We use the algorithm discussed in the previous section to obtain
these exponents. Fig. 6 shows that the H\"{o}lder exponents significantly
increase with the pressure, indicating that the fractal dimension of the
heartbeat interval monotonically decreases with negative pressure. At 
recovery, the fractal properties return to their original values.

Fig. 7 depicts the average H\"{o}lder exponent, indicated by the mean value $%
h_{0}$, and by the standard deviation $\sigma $ for the 12 individuals
against the pressure. As explained in the previous section the standard
deviation measures the variability of the H\"{o}lder exponent values that
can be related to the multifractality of a time series. The data indicates
that the average mean value has a steady increase from $\overline{h_{0}}%
\approx 0.177\pm 0.072$ at $P=0$ mmHg to $\overline{h_{0}}\approx 0.332\pm
0.104$ at $P=-60$ mmHg. The standard deviation is indicated by the error
bars. At the recovery when the negative pressure returns to zero we get $%
\overline{h_{0}}\approx 0.157\pm 0.062$. We also observe that by increasing
the negative pressure the average standard deviation tends to increase. The
values we measure are $\overline{\sigma _{0}}=0.072$, $\overline{\sigma _{15}%
}=0.062$, $\overline{\sigma _{30}}=0.066$, $\overline{\sigma _{45}}=0.077$, $%
\overline{\sigma _{60}}=0.104$ and $\overline{\sigma _{0}}=0.062$ at the
recovery. As anticipated, an increase of the standard deviation of the
H\"{o}lder exponents for equal length time series may be interpreted as an
increase of multifractal properties of the time series. Even if the time
series taken for each value of the pressure are relatively short (approx.
200 data points for $P=0$ mmHg and approx. 300 data points for $P=-60$ mmHg)
and, therefore, the statistics may be not optimum. However, the almost
monotonic increase of the standard deviation with the pressure followed by
an abrupt decrease at the recovery may reasonably indicate a reliable
increase of the multifractality of the heartbeat interval time series with
increasing negative pressure.

To emphasize the properties of heartbeat data at various temporal
resolutions we need to make use of WMA of the data. In Fig. 4 we show the
WMA of the heartbeat time interval depicted in Fig. 1. These data are analyzed
using the MODWT with the LA8 wavelet filter for four levels of resolution,
that is, until the wavelet scale $J=4$. As explained in the previous section the
smooth curve S4 captures the smooth trend of the fluctuation at the scale of 
$\tau =2^{4}=16$ consecutive heartbeats. The details curves D1-D4 capture
the local fluctuations of the signal at each of the four time scale $\tau
=2^{i}$. This means, for example, that D1 captures the fluctuations of the
original signal with a period that is approximately between 2 and 4 beats.
We recall that the shortest possible period that can be detected in a time
series is 2 time units. The detail D2 captures the fluctuations with a
period  in the interval between 4 and 8 beats, and
similarly for the following details. According to Eq. (\ref{decomrel}) the
sum of the four details D1-D4 and the smooth curve S4 gives the original
datasets shown in Fig. 1.

Fig. 4 stresses the great utility of analyzing times series of complex
systems through the WMA via the MODWT. In fact, the wavelet sensibility to
the local changes of the signal allows us to easily extract 
information that is impossible to obtain by using, for example, the Fourier
transform since the latter  averages the changes of the entire signal at each Fourier
transform frequency. For example, Fig. 4 clearly shows that the amplitude of
the local decomposition is not constant but fluctuates in time. In
particular we notice that the amplitude of the fluctuations
captured by the details D3 and D4 does not seem to change systematically with
the pressure. Therefore, we can conclude that the change of pressure do not
influence significantly the slow fluctuations with a period larger than 8
heartbeats.

On the contrary, the details D1 and D2 clearly show a significant reduction
of the amplitude of the fast fluctuation with increasing negative pressure.
At the recovery, the amplitude of the fluctuations returns to a value
similar to the initial one for $P=0$ mmHg. The reduction of the amplitude of
the fast fluctuations manifests itself with the increase of the smoothness of
the heartbeat interval time series that we have already noticed to occur by
increasing the negative pressure, as Fig 1 shows. We have also noticed that
an increase of the smoothness yields an average increase of the local
H\"{o}lder exponents, that are associated with the fractal dimension of a
time series, as explained earlier and shown in Figs. 6 and 7. Fig. 4 shows the
importance of using WMA because we are able to isolate the component of the
signal of the heartbeat interval time series that presents dependence on
the pressure and identifies it with the changes captured by the wavelet
details D1 and D2.

Because each detail curve captures the variation of the signal inside a
frequency or period range, it is possible to also study how the average
period of the fluctuations at a particular wavelet scale are affected by the
pressure. Fig. 8 shows the average mean period measured in number of beats
of the fluctuation associated to each detail curve obtained by the WMA for
12 subjects. We stress that these calculations are made without any kind of interpolation between the data. So the natural unit of the time sequence is the number of heartbeats. To get the results in physical time units we have to convert the results obtained with the heartbeat number in physical time units by using the original R-R sequence or we may attempt a direct calculation of WMA after an interpolation procedure that generates an even time series in a physical units. However,  interpolation alters  data, so we prefer the first procedure. Fig. 9 shows the average mean period measured in physical time
units of the same fluctuations. We decided to plot the results concerning D1, R2, D3 and D4 curves.
We stress that we are considering the residual R2 that is given by $R_2=D_1+D_2$ according to (\ref{decomrelrr}) rather than the D2 curve. We prefer to use the residual R2 instead of the detail D2 because the range from 2 to 8 heartbeats is supposed to capture the high frequency component of the R-R signal that is coupled with the respiratory signal. The residual R2 fully contains such  information that, instead, is split between D1 and D2 curves with D1 being a kind of harmonic of D2.   For this experiment we recall that the subjects breathed in synchrony with a metronome sets  a pace of 15 breaths per minute (0.25 Hz).  So we expect to find such a frequency in the R2 curves. The detail D3 captures  the range from 8 to 16 heartbeats and corresponds to  the low frequency component of the R-R signal. Higher levels of details, that are included in the S3 smooth curve, capture the very-low component of  the interbeat signal.

The results plotted in Figs. 8 and 9
concerning the detail curves D3 and D4 show a more erratic behavior with
large errorbars and their interpretation remains unclear.  Fig. 8 shows an increase of the heartbeat number for cycles captured by   D3 curve with increasing negative pressure. Fig. 9 shows that in physical time unit the period of these D3 cycles remains almost constant at almost 10 seconds, that is, at a frequency of 0.1 Hz, while the period of the cycles associated with D4 detail curve tends to decrease with increasing negative pressure. 

The results concerning the fast fluctuations,  curves D1 and R2, show
more regular patterns. In particular we notice that by increasing the
negative pressure, Fig. 8 shows that the average mean period of the
fluctuations measured in heartbeat number is almost constant for details D1,
D3 and D4. Instead, the results concerning the  R2 residual shows a sensitive
steady increase of the average number of heartbeats from almost 4.1
heartbeats per cycle to 4.9 heartbeats per cycle with increasing negative
pressure from $P=0$ mmHg to $P=-60$ mmHg. At  recovery the average number
of heartbeats returns to  4.1 heartbeat per cycle. By looking at the same
results expressed in physical time units, see Fig. 9, we observe that the
average period of the fluctuations captured by the R2 residual curve is almost
constant with the pressure. We measure  4 seconds that correspond to a frequency of 0.25 Hz. Instead, the
average period of the fluctuations captured by the D1 detail curve shows a
steady decrease from 3.2 to almost 2.1 seconds with increasing negative
pressure.
The different behaviors characterizing the wavelet detail D1 and R2 depicted
in Figs. 8 and 9, suggest that the two curves can be partially decoupled because they respond differently to the
increases in negative pressure.

Fig. 10 shows a comparison between the R2 residual
curve of the heartbeat interval and respiratory signals of a person for 60
seconds under normal conditions. These data were  obtained in a second experiment. Figure 10 shows a very strong correspondence
between the fluctuations captured by the R2 residual curve and the respiratory
cycle. This comparison suggests that the  R2
residual curve has the physiological meaning of describing the fluctuations of respiratory sinus arrhythmia. In fact, as Fig. 9 shows,  the fluctuations captured by the R2 residual  have the frequency of 0.25 Hz of the respiratory signal that is kept constant during the hypovolemia experiment.

\section{Summary and conclusions}

The analysis of the data suggests that progressive central hypovolemia induced with LBNP:

1) Causes decreases of R-R intervals and, therefore, increases of heartbeat rate.

2) Decreases heart rate variability and therefore smoothes out the heartbeat time series.

3) 	Reveals that two measures of the decrease in the heartbeat rate variability with
increasing negative pressure are an increase in the H\"{o}lder exponent and
a decrease in the fractal dimension of the time series.

4)	Reveals that accompanying the increase in the H\"{o}lder exponent, with increasing
negative pressure, is an increase in the multifractality of the time series
as measured by the width of the H\"{o}lder exponent distribution. This
non-stationary aspect of the data was also observed in the study of human
gait data, when the walkers were put under physical stress by asking them to walk faster or slower than normal \cite{scafetta1,scafetta2}.

5) 	Reveals that independent measures of the heartbeat rate response to the increases in
negative pressure are the wavelet transform details D1 and D2. Both these
non-periodic functions display a monotonic decrease in amplitude with
increasing negative pressure.

6) 	Suggests that the amplitude of the  higher-order details, D3 and D4, are independent of the
changes in the negative pressure. However, the period of the cycles captured by D4 tends to decrease with increasing negative pressure. This suggests that part of the spectral energy of the very-low frequency ($f<0.04$ Hz) component of the RR signal moves toward the low frequency ($0.04<f<0.15$ Hz) component of the same signal. 

7) 	Suggests that the wavelet transform residual curve R2 is driven by the respiratory
cycle. The residual R2 captures the high frequency component of the RR signal ( $0.15<f<0.4$ Hz). This is an indication that the wavelet transform is able to sort
through the fluctuations of the cardiac time series and highlight particular
physiological phenomena.

These separate observations made from the analysis of the heartbeat time
series stressed by central hypovolemia are interesting in themselves, but they
suggest a much more significant finding. They suggest that the influence of stress on a physiological system is manifest in the
non-stationary character of the time series and that the degree of
non-stationarity can be measured through the level of fractality and multifractality of the
 time series regarding the stressed system.\\

 {\bf Acknowledgment:}  
This work was supported in part by a 
Faculty Research Fellowship through the US Army Institute of Surgical 
Research under the auspices of the US Army Research Office Scientific 
Services Program administered by Battelle (DAAH04-96-8806).
We acknowledge Dr. Victor Convertino, Institute of Surgical Research, Ft. Sam Houston TX, as having contributed the LBNP data.
We thank the research group of prof. B Ghelarducci, Dipartimento di Fisiologia e Biochimica, Universit\'a di Pisa, and 
Dr.  M. Varanini, Institute of Clinical Physiology, CNR-Pisa, for the respiratory and contemporary heartbeat signal used in the second experiment. 
N.S. thanks the Army Research Office for support under grant DAAG5598D0002.

\newpage 
\begin{figure*}[tbp]
\epsfig{file=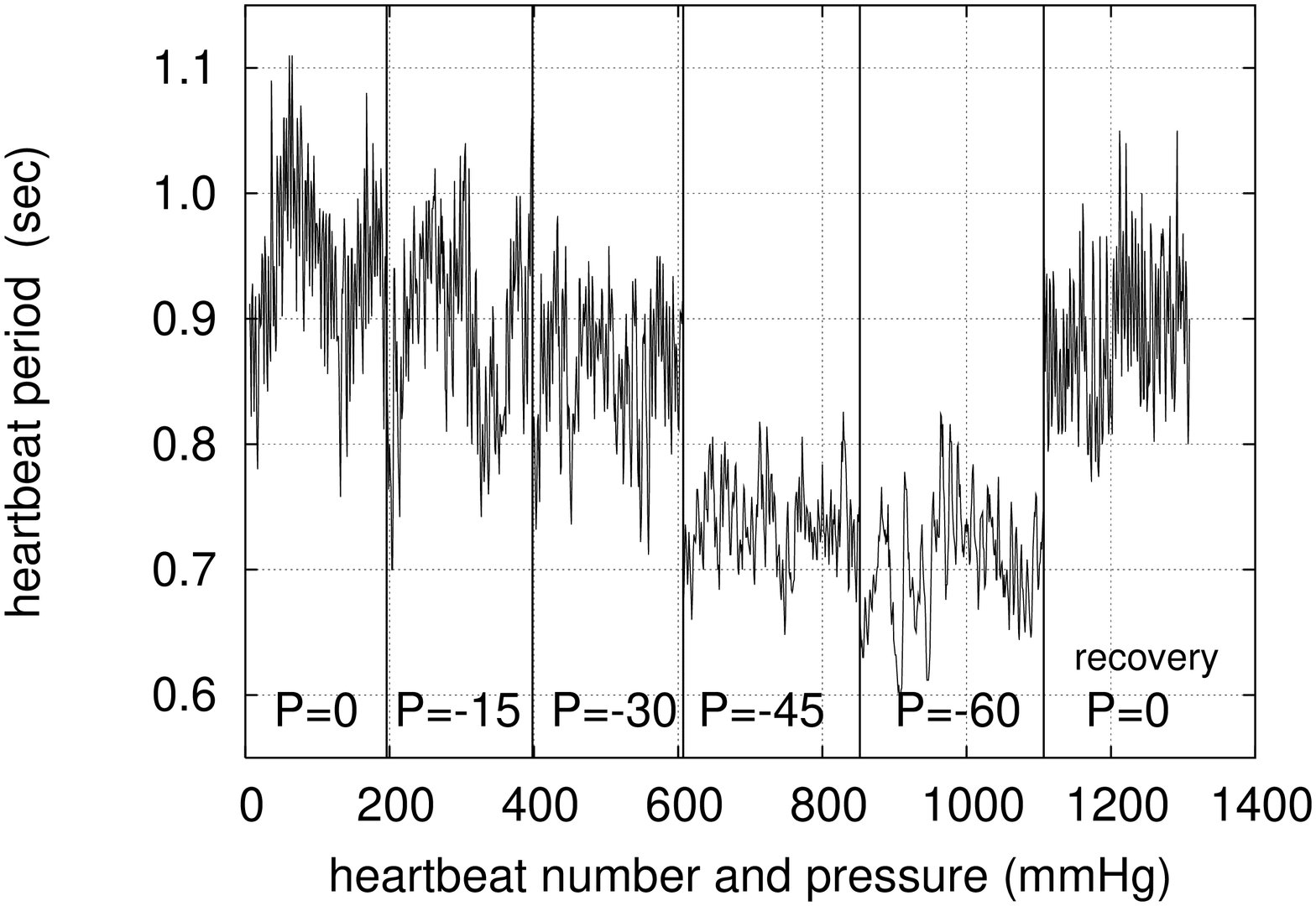,height=13cm,width=10cm,angle=-90}
\caption{Typical response of the heartbeat period at pressure change. }
\end{figure*}

\begin{figure*}[tbp]
\epsfig{file=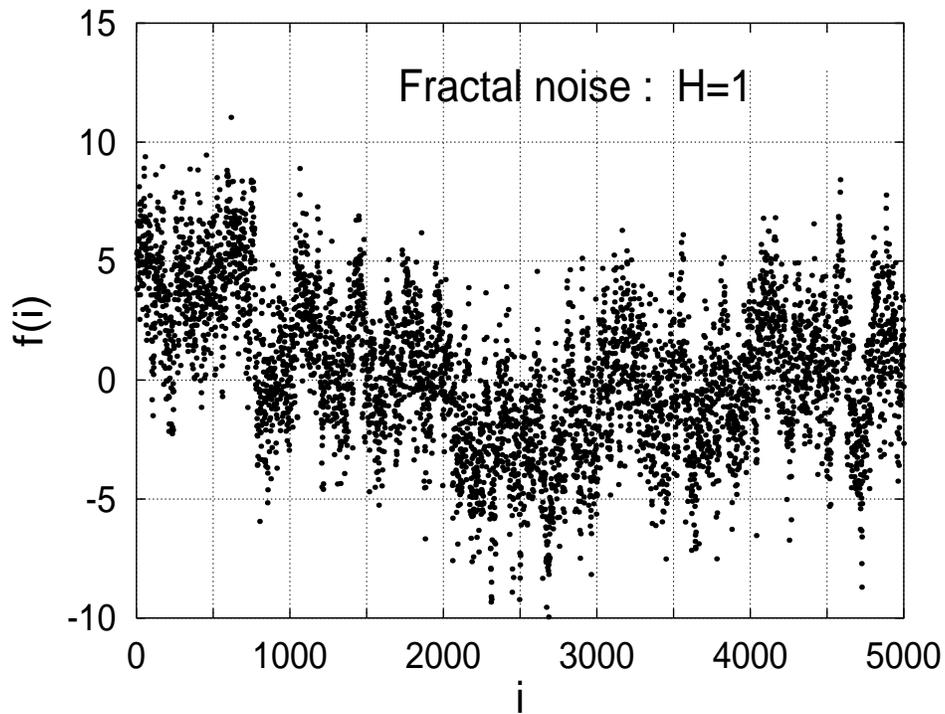,height=13cm,width=10cm,angle=-90}
\caption{ Computer-generated fractal noise (FGN) with Hurst coefficient $H=1$%
, also known as $1/f$ noise or \textit{pink} noise is shown. }
\end{figure*}

\begin{figure*}[tbp]
\epsfig{file=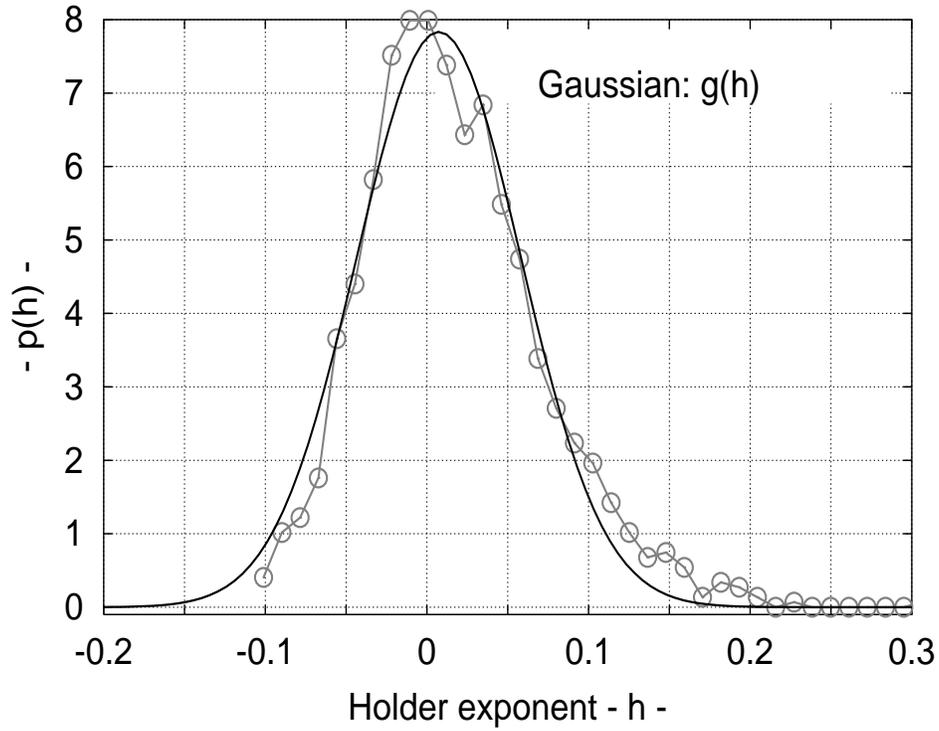,height=13cm,width=10cm,angle=-90}
\caption{ Histogram and probability density estimation of the H\"older
exponents for the computer-generated fractal noise with Hurst coefficient $%
H=1$ shown in Fig. 2. The fitting curve is a Gaussian (\ref{gaussf})
centered in $h_0=0.007 \pm 0.001$ and with a width $\sigma=0.051 \pm 0.001$.
The mean H\"older exponent given by Eq. (\ref{linfitesa}) is $\overline{h}%
=-0.004 \pm 0.008$ and correspond to the maximum of the distribution. }
\end{figure*}

\begin{figure*}[tbp]
\epsfig{file=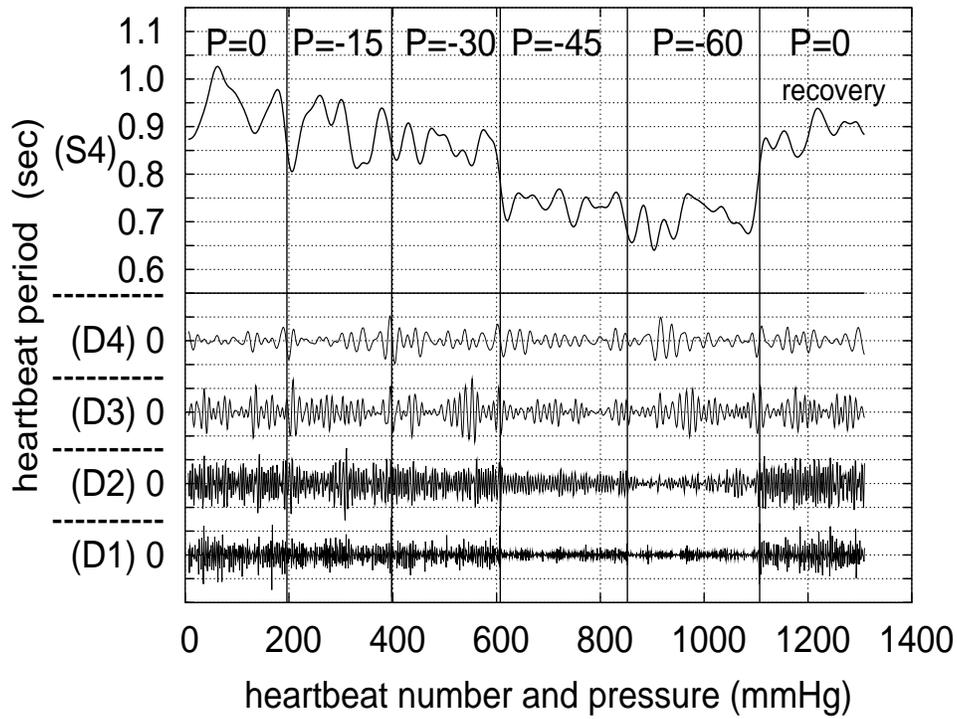,height=13cm,width=10cm,angle=-90}
\caption{ Wavelet multiresolution analysis for $J=4$ for the data shown in
Fig. 1. }
\end{figure*}

\begin{figure*}[tbp]
\epsfig{file=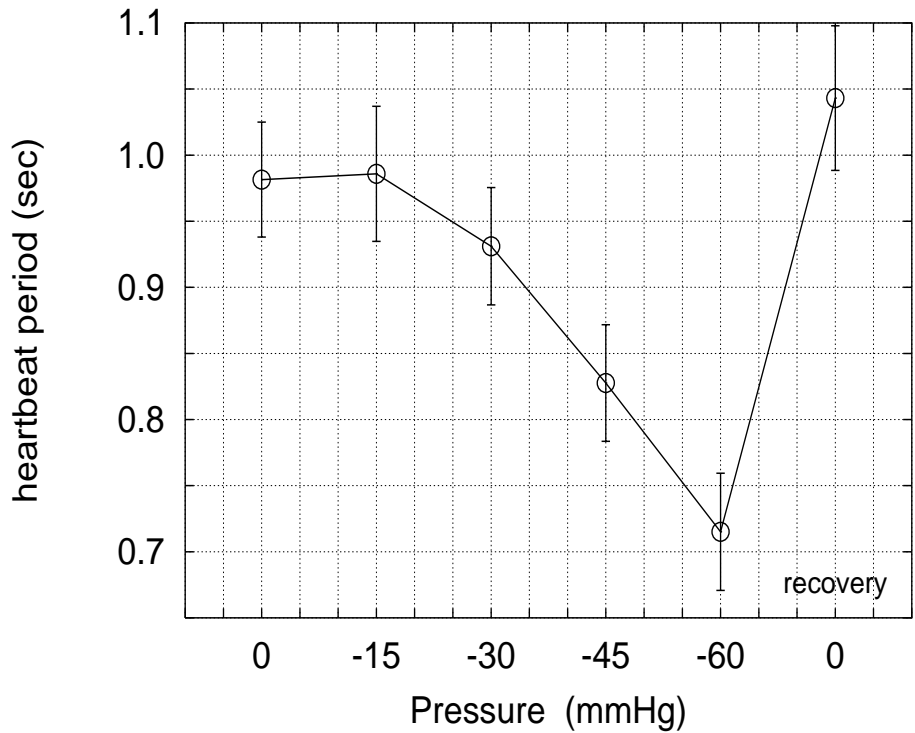,height=13cm,width=10cm,angle=-90}
\caption{ Average heartbeat period as function of the pressure for the 12
datasets. }
\end{figure*}

\begin{figure*}[tbp]
\epsfig{file=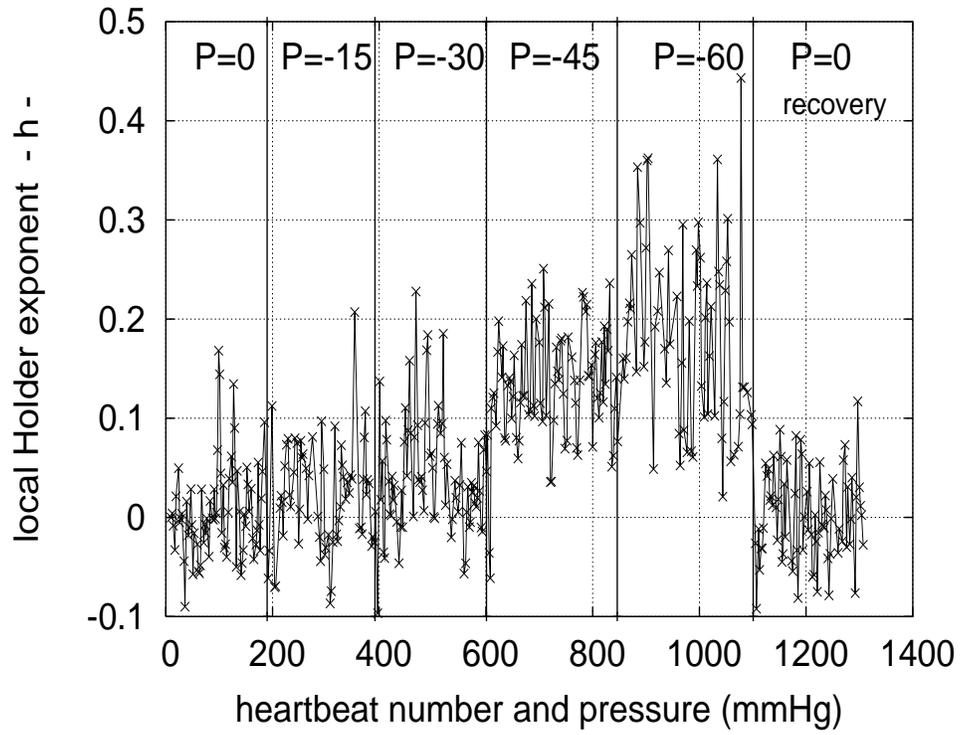,height=13cm,width=10cm,angle=-90}
\caption{ Local H\"older exponent for the singularities of the hearbeat
interval time series shown in Fig. 1. }
\end{figure*}

\begin{figure*}[tbp]
\epsfig{file=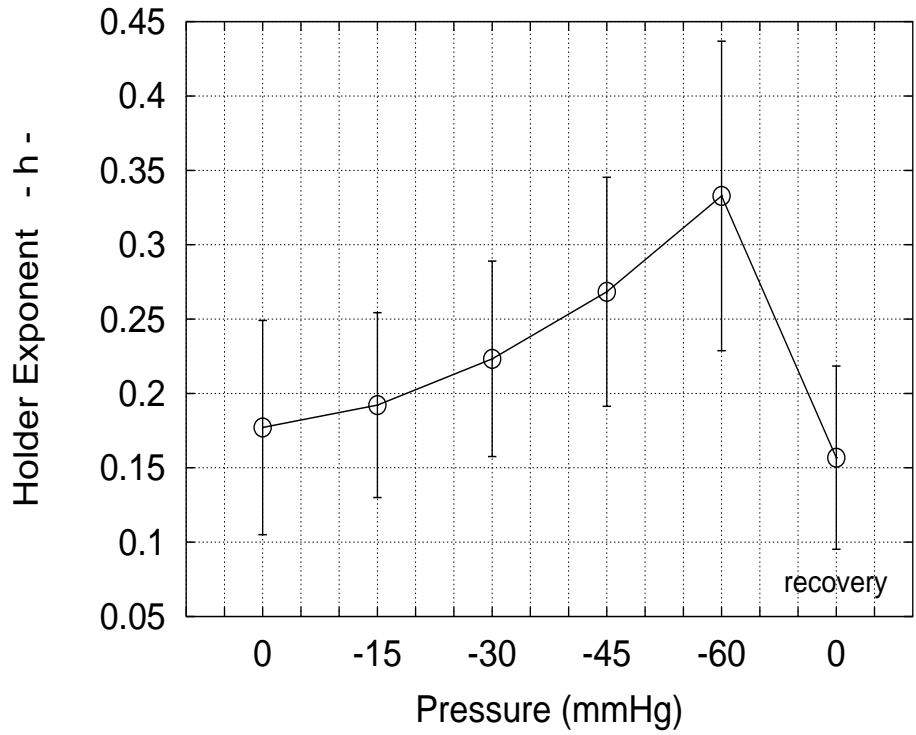,height=13cm,width=10cm,angle=-90}
\caption{ H\"older exponent distribution for the heartbeat period time
series as function of the pressure. The error bars indicate the average
standard deviation of the H\"older exponent distribution for the 12
datasets. }
\end{figure*}

\begin{figure*}[tbp]
\epsfig{file=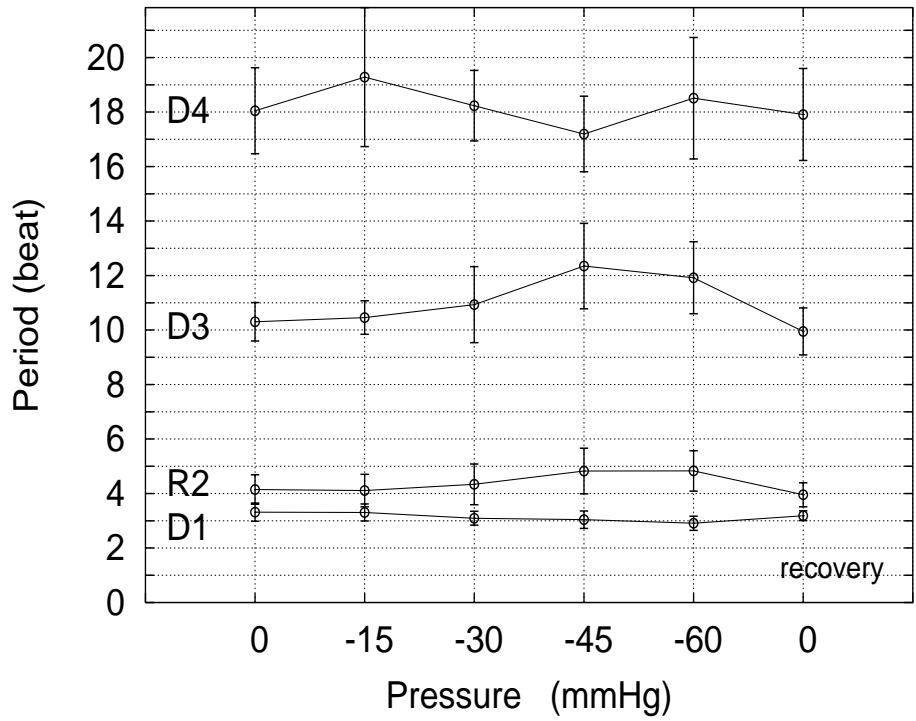,height=13cm,width=10cm,angle=-90}
\caption{ Average mean period measured in number of heartbeats of the
fluctuation associated to D1, R2, D3, and D4 detail curve obtained by the WMA for $J=4$
for the 12 datasets. }
\end{figure*}

\begin{figure*}[tbp]
\epsfig{file=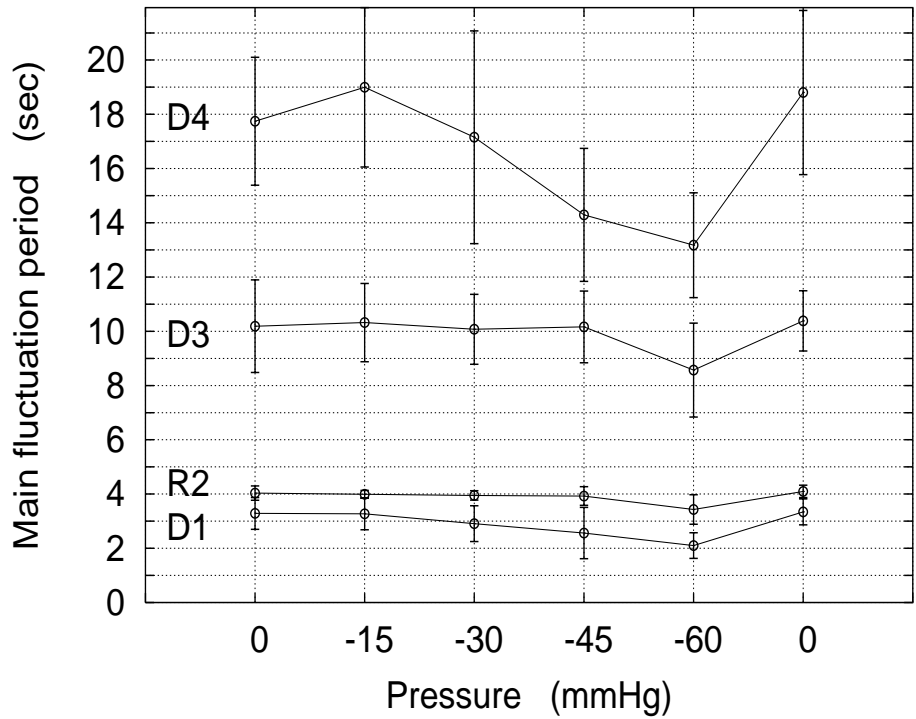,height=13cm,width=10cm,angle=-90}
\caption{ Average mean period measured in physical time units of the
fluctuation associated to D1, R2, D3, and D4 curve obtained by the WMA for $J=4$
for the 12 datasets. Note that the period associated to the residual R2 remains 4 second under different intensity of negative pressure. Four seconds is  the period of the breathing cycle. }
\end{figure*}

\begin{figure*}[tbp]
\epsfig{file=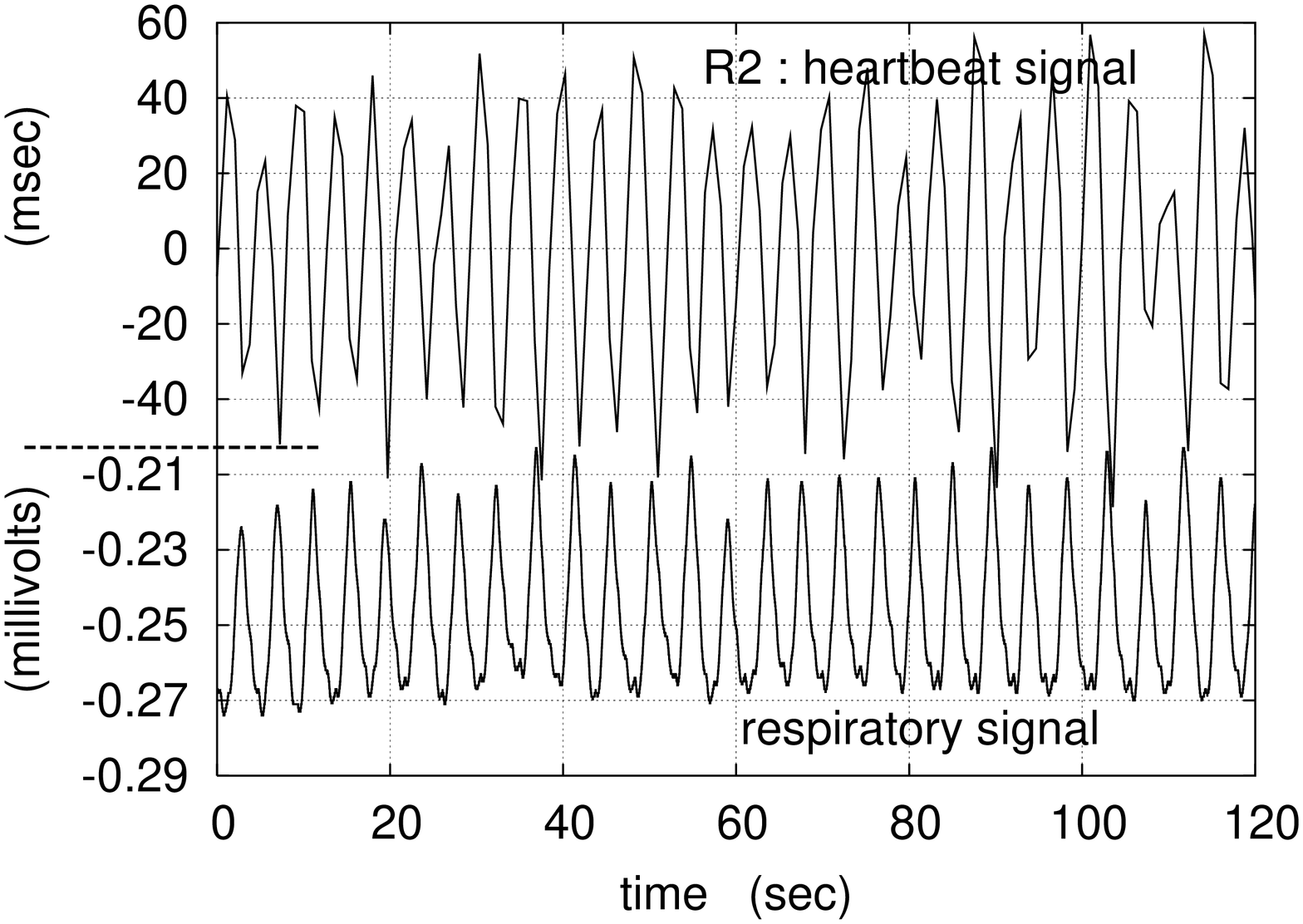,height=13cm,width=10cm,angle=-90}
\caption{ Comparison between R2 detail curve of the heartbeat interval signal and the contemporary
respiratory signal of a subject. }
\end{figure*}


\begin{thebibliography}{99}
\bibitem{bass94}  J.B. Bassigthwaighte, L.S. Liebowitch and B.J. West, 
\textit{Fractal Physiology}, Oxford University Press, Oxford (1994).

\bibitem{feders}  J. Feder, \emph{Fractals}, Plenum Publishers, New York
(1988).

\bibitem{ivanov99}  P. C. Ivanov, M. G. Rosenblum, L. A. Nunes Amaral, Z. R.
Struzik, S. Havlin, A. L. Goldberger and H. E. Stanley, ``Multifractality in human heartbeat dynamics," \textit{Nature} 
\textbf{399}, 461-465 (1999).

\bibitem{latka02}  B. J. West, M. Latka, M.  Glaubic-Latka, ``Multifractality of cerebral blood flow,"
Physica A 318 (3-4): 453-460  (2003).



\bibitem{scafetta1}  N. Scafetta, L. Griffin and B. J. West, ``Holder exponent spectra for human gait," in press on
Physica A (2003).

\bibitem{scafetta2} B. J. West and N. Scafetta, ``A non linear model for human gait," in press on Phys. Rev. E (2003).

\bibitem{Covertino}  V. A. Convertino. ``Lower body negative pressure as a tool for research in aerospace physiology and military medicine." Journal of Gravitational Physiology, 8:1-14, (2001).

\bibitem{Akselrod} S. Akselrod, D. Gordon, F.A. Ubel,  D.C. Shannon,  A.C. Barger,  and R.J. Cohen,  ``Power spectrum analysis of heart rate fluctuation: a quantitative probe of beat-to-beat cardiovascular control." Science, 213:220-222, (1981).

\bibitem{Malliani}  A. Malliani, ``The pattern of sympathovagal balance explored in the frequency domain." News in Physiological Sciences, 14:111-117, (1999).

\bibitem{Brown}  T. E. Brown, L.A. Beightol,  J. Koh,  and D.L. Eckberg.  ``Important influence of respiration on human R-R interval power spectra is largely ignored." Journal of Applied Physiology, 75:2310-2317, (1993).

\bibitem{struzik}  Z. R. Struzik, ``Determining local singularity strengths and their spectra with the wavelet transform," Fractals, Vol. 8, No. \textbf{2}, 163-179
(2000).

\bibitem{percival}  D. B. Percival and A. T. Walden, \textit{Wavelet Methods
for Time Series Analysis}, Cambridge University Press, Cambridge (2000).

\bibitem{2Mandelbrot}  B.B. Mandelbrot, \textit{The Fractal Geometry of
Nature}, Freeman, New York, (1983).

\bibitem{politi} R. Badii and A. Politi, {\it Complexity, Hierarchical structures and scaling in physics}, Cambrige University Press, UK 1997. 

\bibitem{daubechies}  I. Daubechies, \textit{Ten Lectures On Wavelets}, SIAM
(1992).

\bibitem{Mallat}  S. G. Mallat, \textit{A Wavelet Tour of Signal Processing}
(2nd edition), Academic Press, Cambridge (1999).



\bibitem{mallat2}  S. G. Mallat, W. L. Hwang, ``Singularity detection and processing with wavelets",  IEEE Trans. on Information
Theory \textbf{38}, 617-643 (1992).

\bibitem{mallat3}  S. G. Mallat, S. Zhong, ``Complete Signal Representation with Multiscale Edges" IEEE Trans. PAMI \textbf{14}, 710-732 (1992).

\bibitem{arneodo1}  J. F. Muzy, E. Bacry,  . Arneodo,  ``Multifractal formalism for fractal signals - The structure-function approach versus the wavelet-transform modulus-maxima method," Phys. Rev. E \textbf{47}, No. 2, 875-884 (1993).

\bibitem{arneodo2}  A. Arneodo, E. Bacry, J. F. Muzy, ``The multifractal formalism revisited with
wavelets," \textit{International
Journal of Bifurcation and Chaos} \textbf{4}, No. 2, 245-302 (1994).

\bibitem{struzik2}  Z. R. Struzik,  ``Removing Divergences in the Negative Moments of the multi-fractal Partition Function with the Wavelet Transform," CWI Report, INS-R9803 (1998).





\end{thebibliography}
\end{document}